\documentstyle[fancyheadings,prl,
amsmath,amsfonts,isolatin1,aps,prc,epsf,amssymb,epsfig,multicol]{revtex}

\begin{document} 
 
\sloppy 
 
\draft 
 
\bibliographystyle{srt}
 
\title{ Constraints on a scalar-pseudoscalar Higgs mixing at future 
$e^+ e^-$ colliders: an update} 
 
\author{ 
A. Chalov$^{a}$ ,  
A. Likhoded$^{a}$\footnote{andre@mx.ihep.su}  and    
R. Rosenfeld$^{b,}$\footnote{rosenfel@ift.unesp.br}}

 \address{\it  $^a$ Institute of High Energy Physics} 
\address{\it Protvino, Moscow Region, Russia} 
\vspace{1cm}

\address{\it  $^b$ Instituto de F\'{\i}sica Te\'orica - UNESP} 
\address{\it Rua Pamplona, 145} 
\address{\it 01405-900 S\~{a}o Paulo, SP, Brazil}

\maketitle 
 
\vspace{0.1cm} 
 
\begin{abstract} 
We perform an update of our previous analysis of the constraints on possible 
deviations of $H b \bar{b}$ coupling parametrized as 
$\frac{m_b}{v} (a + i \gamma_5 b) $,
arising from a scalar-pseudoscalar mixing, where the process  
$ e^+ e^- \to b \bar{b} \nu \bar{\nu}$ was used.
In this paper we include a complete simulation of
the process $ e^+ e^- \to b \bar{b}e^+ e^-$ and combine these results to 
obtain tighter bounds on the deviations of the parameters $a$ and $b$ from 
their Standard  Model values that could be measured at the Next Linear
Collider.
\end{abstract} 
 
\vspace{0.2cm} 
 
\noindent 
PACS Categories:    98.80.Cq 
 
\vspace{0.2cm}

 
\begin{multicols}{2} 
\relax 
 
\section{Introduction} 
 
The origin of fermion masses and mixings is an outstanding open problem in 
particle physics. In the Standard Model (SM), the Higgs mechanism is  
responsible 
for the electroweak symmetry breaking and mass generation via {\it ad-hoc} 
Yukawa couplings. There are reasons to believe that the SM is not the final
model and a complete study of the coupling of the 
lightest boson, which we will call the Higgs boson, 
to fermions can provide hints on new physics beyond the SM.

In a recent letter \cite{PRL}, we performed a realistic simulation of the process 
$ e^+ e^- \to b \bar{b} \nu 
\bar{\nu}$, where $\nu$ can be an electron, muon or tau neutrino, in the
environment of a future Linear Collider with a center-of-mass energy of 
$\sqrt{s} = 500$ GeV with an accumulated luminosity of 1 ab$^{-1}$, based 
on the TESLA design \cite{Tesla}.
In  particular, we noticed that weak gauge boson fusion is the dominant 
contribution to the subset 
of diagrams containing the Higgs boson 
for $M_H < 180$ GeV at $\sqrt{s} \ge 500$ GeV and hence this process is
sensitive to the Higgs boson couplings to $b$-quarks. 
We have used a similar technique to investigated the possibility of detecting 
deviations 
from the SM in the Higgs couplings to $\tau$-leptons  at future $e^+ 
e^-$ colliders \cite{us}, which can be improved by using $\tau$ spin
correlations \cite{tauspin}.

In this Brief Report, we update the result \cite{PRL} by combining it with 
the results
of a detailed analysis of the process $ e^+ e^- \to b \bar{b}e^+ e^-$.
This process has a very clean final state, with no missing energy and easy
reconstruction, which can compensate for its smaller rate. It certainly
must be included in a global analysis of the $H b \bar{b}$ vertex.
 
In order to perform our analysis, we will assume 
that the Higgs boson has already been discovered at the Large Hadron Collider
and  concentrate on the determination of its coupling to $b$-quarks. 
 In extensions of the SM with extra scalars and pseudoscalars, the lightest 
spin-0 particle can be an admixture of states without a definite parity.  
Hence, we parametrize the general $H b \bar{b}$ coupling as: 
\begin{equation} 
\frac{m_b}{v} (a + i \gamma_5 b)\;,  
\end{equation} 
where $v = 246$ GeV, $m_b$ is the $b$-quark mass and $a = 1$, $b = 0$  in the  
SM.  In CP-violating extensions of the SM, deviations of these parameters may be
generated at tree-level \cite{CPstudies}.

We will present results  
considering $a$ and $b$ as independent parameters and also for the cases of  
fixed $a=1$, free $b$ and fixed $b=0$, free $a$. There is a 
region of insensivity around circles in the $a-b$ plane 
since we can't at this level of analysis disentangle the effects of 
$a$ and $b$.

In the SM the $ e^+ e^- \to b \bar{b} e^+ e^- $ process is determined by 50
Feynman diagrams. Only 2 of these (Higgs radiative production, $Z^\ast \to H(\to
b \bar{b}) Z$,
and vector boson fusion, $ Z^\ast Z^\ast \to H(\to
b \bar{b})$) can be considered as signal,
where deviations from SM Higgs couplings to $b$-quarks show up. The remaining
diagrams are not changed by new physics in the Higgs sector.

The total SM cross section for the process $ e^+ e^- \to b \bar{b} e^+ e^- $ 
is approximately $43$ fb for $M_H = 120$ GeV  
at $\sqrt{s} = 500$ GeV, with cuts in the scattering angle between initial 
and final electrons$, |\cos \theta_{ee}| \leq 0.9962$, and in the final 
electron-positron invariant mass, $M_{e^+ e^-} > 2$ GeV. The angular cut avoids
the region of $\pm 5^0$ around the pipeline, used by the beam pipe itself and
by the Small Angle Monitor (SAM) for luminosity measurements. The invariant mass
cut insures that the electrons and positrons will be detected in the lead glass
calorimeter \cite{Tesla}.
Backgrounds from $e^+ e^- \to e^+ e^- Z Z \to e^+ e^- b \bar{b} \nu \bar{\nu}$,
$e^+ e^- \to Z Z Z \to e^+ e^- b \bar{b} \nu \bar{\nu}$, etc, can be reduced 
to the $0.1$ fb level by appropriate cuts\cite{lcnote}.

In our analysis we will assume SM couplings of 
the Higgs to the electroweak gauge bosons. 
A full analysis taking into account non-SM HWW and HZZ couplings would 
introduce many more parameters and is beyond the scope of this paper.
 
 
\section{Analysis and Results} 
 
We performed our Monte Carlo simulation by generating observables 
represented as series in the $a$ and $b$ couplings multiplied by kinematical 
factors:  
\begin{eqnarray} 
\frac{d\sigma }{d {\cal O} }  & = & A_0 + a\cdot A_1  
+ a^2\cdot A_2+ ab\cdot A_3+  \nonumber  \\
& &   b\cdot A_4+ b^2\cdot A_5 \mbox{\dots } 
\end{eqnarray} 
where ${\cal O}$ is any observable and the $A_i$ terms are purely kinematical  
structures which do not contain 
any $a$ and $b$ dependence and results from the amplitude squaring and phase 
space integration. In this way we only simulate the process once for each
observable. In our case, terms linear in $b$ parameter vanish and hence
$A_3 =  A_4 = 0$. We have studies the following observables:
transverse $b$-quark momentum $p_{Tb}$, $b \bar{b}$ invariant mass
$M_{b \bar{b}}$,  $\cos\theta_{eb}$, where $\theta_{eb}$
is the scattering angle between the $b$-jet and initial beam directions, and
the $T$ correlation,
defined by $T = \frac{1}{(\sqrt{s})^3}
\vec{p}_{el} \cdot    (\vec{p}_{b} \times \vec{p}_{\bar{b}})$.
 
The event sample reproducing the expected statistics at TESLA was generated 
using our Monte Carlo package while the detector response was simulated with 
the code {\tt SIMDET} version 3.01\cite{simdet}. We assume an efficiency  
for $b-$jet pair reconstruction of $ \varepsilon_{bb}=56$ \%,  
which is based on the $b$-tag 
algorithm, as assumed in ref. \cite{lcnote}. In our simulations we used $M_H = 
120$ GeV.

\begin{center} 
\begin{figure} 
\hskip-1.0cm  
\centerline{\epsfxsize=1\hsize \epsffile{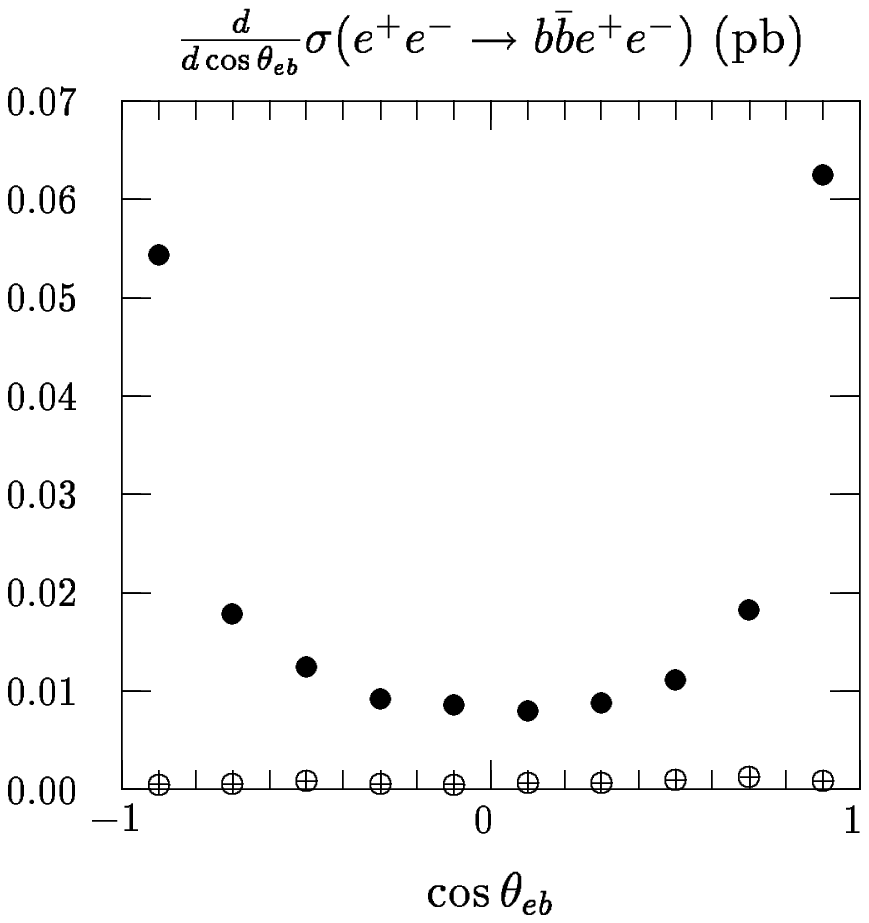} 
\hskip-4.5cm  
\epsfxsize=1\hsize \epsffile{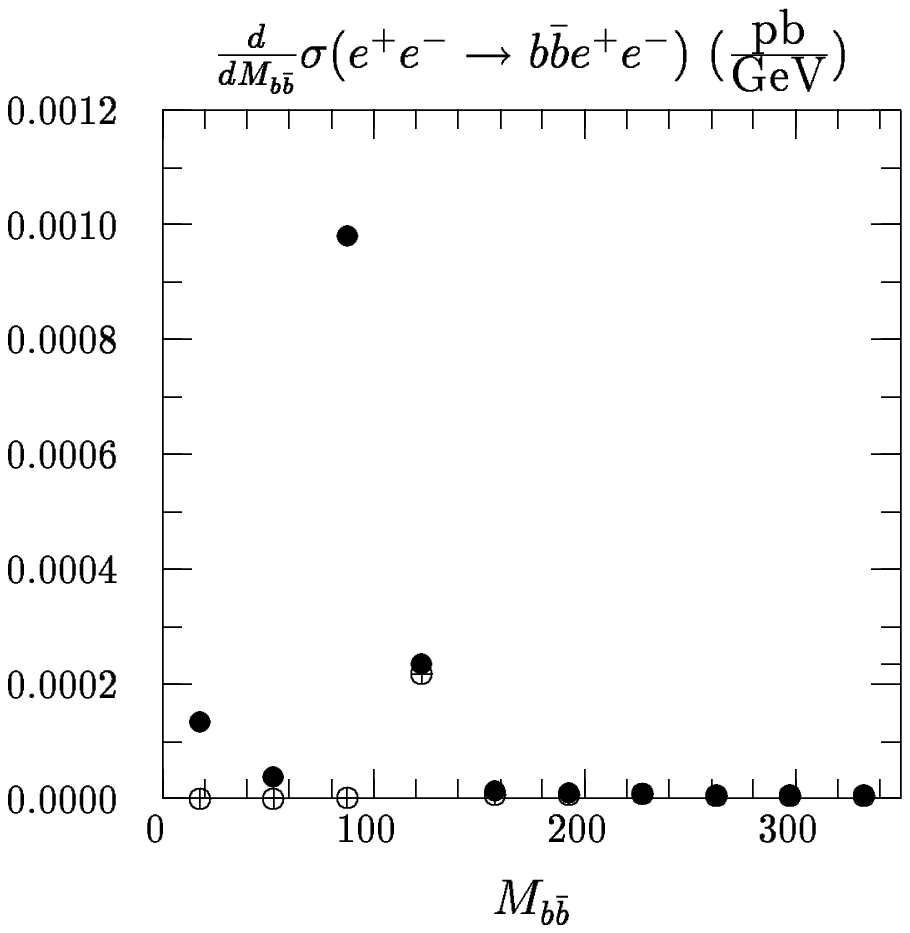}} 
\vspace*{-5cm} 
\caption{  
The differential cross sections for $\cos\theta_{eb}$ and
$M_{b \bar{b}}$   
for the process $e^+e^-\to b \bar{b}e^+e^- $.  
Solid circles are the full  
SM result while crossed circles are the contribution from Higgs only  
(including interference effects). } 
\label{distributions} 
\end{figure} 
\end{center}

In Figure \ref{distributions} we show, for comparison purposes, the  
differential distributions in the invariant $b \bar{b}$ mass and  
$\cos\theta_{eb}$,
for the total SM contribution and for the Higgs contribution only  
(including interference with SM). Notice the $Z$-boson peak  and the smaller
Higgs peak in the $M_{b \bar{b}}$ distribution, and how the Higgs
contribution dominates around its peak.

\vspace*{-4.5cm} 
\vspace*{-2.7cm}
\begin{figure} 
\centerline{\epsfxsize=1\hsize \epsffile{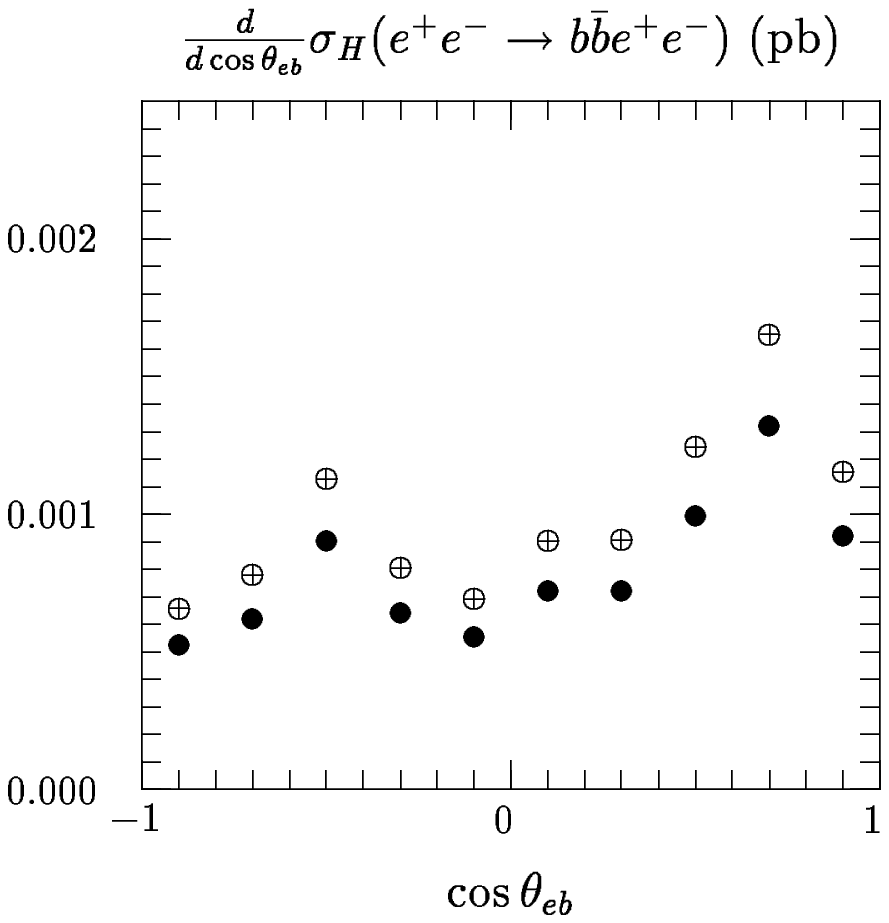}} 
\vskip-5.5cm  
\caption{ 
Contribution of Higgs diagrams to the differential $\cos\theta_{eb}$  
distribution in the  $e^+e^-\to b \bar{b} e^+e^-$ process for 
Standard Model ($a=1, b=0$, black dots) and $a=1.0, b=0.5$ (crossed dots).} 
\label{ab} 
\end{figure} 

In order to demonstrate the effect of different values of the parameters $a$ and 
$b$, we show in Figure \ref{ab} the $\cos\theta_{eb}$ distribution arising from
the Higgs contribution for the SM  
($a=1$, $b=0$) 
compared to the case with $a = 1$, $b = 0.5$.  
We see that the shapes are very similar, as expected, but  
the levels can be noticeable different.

Another important aspect is the assumption about the detector performance  
and possible sources of the systematic uncertaintes. We  
include the anticipated  systematic errors of 0.5\% in the luminosity  
measurement, 
1\%  in the acceptance determination, 1\% in the branching ratios, and 
1\% in the background substraction, and assume the Gaussian nature of the  
sytematics \cite{Tesla}. To place bounds on the $H b \bar{b}$ couplings, we  use a standard  
$\chi^2$-criterion to analyse the events. After various kinematical
distributions were examined, we found that the  
most strict bounds are achieved from $\cos\theta_{eb}$ distribution
by dividing the distribution event samples into 10 bins.  
The experimental error $\Delta 
\sigma^{exp}_i$ for the $i^{th}$ bin is given by: 
\begin{equation} 
\Delta\sigma^{exp}_i = \sigma^{SM}_i \sqrt{\delta^2_{syst} + \delta^2_{stat} } 
\end{equation} 
where  
\begin{equation} 
\delta_{stat}  = \frac{1}{\sqrt{\sigma^{SM}_i \varepsilon_{bb} \int  
{\cal L} dt  
}} 
\end{equation} 
and $\delta_{syst}^2$ is the sum in quadrature of the systematic uncertainties 
mentioned above.

Below we present our final results for a TESLA-like 
environment \cite{Tesla} with a center-of-mass energy of 500 GeV 
and for $M_H= 120$ GeV. 
We investigated three possible scenarios for the luminosities: 100 fb$^{-1}$,
1 ab$^{-1}$ and 10 ab$^{-1}$.

The bounds that can be obtained at 95\% confidence level from
$ e^+ e^- \to b \bar{b} e^+ e^- $ process on the 
$\Delta a = a-1$ and $b$ parameters are: 
\begin{eqnarray} 
-0.09 &\leq& \Delta a\leq 0.08\; \mbox{ for ${\cal L} = 100$ fb$^{-1}$}; \\ 
\nonumber 
-0.056 &\leq& \Delta a\leq 0.055\; \mbox{ for ${\cal L} = 1$ ab$^{-1}$}; \\ 
-0.05 &\leq& \Delta a\leq 0.05\; \mbox{ for ${\cal L} = 10$ ab$^{-1}$}, \nonumber 
\end{eqnarray} 
for the case of $b=0$ and free $\Delta a$ and 
\begin{eqnarray} 
-0.42 &\leq& b\leq 0.42\; \mbox{ for ${\cal L} = 100$ fb$^{-1}$}; \\\nonumber 
-0.32 &\leq& b\leq 0.32\; \mbox{ for ${\cal L} = 1$ ab$^{-1}$}; \\ 
-0.3 &\leq& b\leq 0.3\; \mbox{ for ${\cal L} = 10$ ab$^{-1}$},\nonumber 
\end{eqnarray} 
for the case of $\Delta a=0$ and free $b$. 
 
We will combine these limits with our previous bounds from the
$ e^+ e^- \to b \bar{b} \nu \bar{\nu}$ process according to the following
procedure. Given two bounds on the same quantitity $X$, say $|X| \leq c_1$ and 
 $|X| \leq c_2$, the combined bound will be 
\begin{equation}
|X| \leq c_3  \;\;\;\ \mbox{where} \; c_3 =  \sqrt{(c_1 \omega_1)^2 +
(c_2 \omega_2)^2},
\end{equation}
and the statistical weights $\omega_1$ and $\omega_2$ are given by
\begin{equation}
\omega_1 = \frac{N_1}{N_1 + N_2} \;\;\;\ \mbox{and} \;\;\;
\omega_2 = \frac{N_2}{N_1 + N_2},
\end{equation}
where $N_1$ and $N_2$ are the number of events for each process.

With these prescription we obtain the improved bounds:
\begin{eqnarray} 
-0.037 &\leq& \Delta a\leq 0.035\; \mbox{ for ${\cal L} = 100$ fb$^{-1}$}; \\ 
\nonumber 
-0.024 &\leq& \Delta a\leq 0.024\; \mbox{ for ${\cal L} = 1$ ab$^{-1}$}; \\ 
-0.022 &\leq& \Delta a\leq 0.022\; \mbox{ for ${\cal L} = 10$ ab$^{-1}$}, \nonumber 
\end{eqnarray} 
for the case of $b=0$ and free $\Delta a$ and 
\begin{eqnarray} 
-0.24 &\leq& b\leq 0.24\; \mbox{ for ${\cal L} = 100$ fb$^{-1}$}; \\\nonumber 
-0.20 &\leq& b\leq 0.20\; \mbox{ for ${\cal L} = 1$ ab$^{-1}$}; \\ 
-0.19 &\leq& b\leq 0.19\; \mbox{ for ${\cal L} = 10$ ab$^{-1}$},\nonumber 
\end{eqnarray} 
for the case of $\Delta a = 0$ and free $b$.

These results can be roughly scaled for moderate variations in the Higgs boson  
mass around $120$ GeV by multiplying the bounds by a factor $(M_H/120 \mbox{ 
GeV})^2$.

\section{Conclusions} 
 
We have performed an update in our previous constraints on deviations of the 
$H b \bar{b}$ coupling from its SM value by including a complete analysis 
of the sensitivity 
due to the process $ e^+ e^- \to b \bar{b}e^+ e^- $ at the next generation 
of linear colliders. These deviations are 
predicted by many extensions of the Standard Model. We improved our previous
bounds by roughly $10$\%.

We showed that future $ e^+ e^-$ linear collider
experiments will be able to probe deviations of $H b \bar{b}$ coupling. 
The weak gauge boson fusion process is instrumental for achieving such a 
precision.  
For a TESLA-like environment, we are able to constrain the couplings at the 
level of a few percent for the $a$ parameter (for fixed $b$) and tens of  
percent for the $b$ parameter (for fixed $a$). These results are comparable to 
the study performed in \cite{Tesla}, where a global fit analysis for ${\cal L} 
= 500$ fb$^{-1}$ and $\sqrt{s} = 500$ GeV has resulted in a relative accuracy of 
$2.2$\% in the $g_{Hbb}$ Yukawa coupling. 
For comparison, the top quark Yukawa coupling can be determined with a 
statistical accuracy of 16\% at the LHC for $M_H = 130$ GeV \cite{LHC}.

In our analysis we can not disentangle the contributions from deviations in the
$a$ and $b$ parameters. However, notice that the  partial width 
$\Gamma_{H\to b \bar b}$ is proportional to $(a^2 +b^2)$. 
Our distributions have a different dependence
\begin{displaymath} 
\frac{d\sigma }{d {\cal O} } = A_0 + a\cdot A_1  
+ a^2\cdot A_2+ b^2\cdot A_3\;.  
\end{displaymath} 
Therefore, if  an independent  measurement 
of $\Gamma_{H\to b \bar b}$ is obtained (for instance, 
from on-mass-shell Higgs production in Higgs-strahlung or in a muonic collider),
one would be able to separate the
$a$ and $b$ contributions and obtain an explicit indication  
of CP violation in the Higgs sector.

\section*{Acknowledgments} 
 
The work of A. Chalov and A. Likhoded is supported 
by the Russian Foundation for Basic Research, grants SS-1303.2003.2 and
04-02-17530, and the Russian Education Ministry grant E02-3.1-96. 
R. Rosenfeld would like to thank CNPq for partial 
financial support. The authors would like to thank A. Belyaev for
valuable remarks.

\def \arnps#1#2#3{Ann.\ Rev.\ Nucl.\ Part.\ Sci.\ {\bf#1} (#3) #2} 
\def \art{and references therein} 
\def \cmts#1#2#3{Comments on Nucl.\ Part.\ Phys.\ {\bf#1} (#3) #2} 
\def \cn{Collaboration} 
\def \cp89{{\it CP Violation,} edited by C. Jarlskog (World Scientific, 
Singapore, 1989)} 
\def \econf#1#2#3{Electronic Conference Proceedings {\bf#1}, #2 (#3)} 
\def \efi{Enrico Fermi Institute Report No.\ } 
\def \epjc#1#2#3{Eur.\ Phys.\ J. C {\bf#1} (#3) #2} 
\def \f79{{\it Proceedings of the 1979 International Symposium on Lepton and 
Photon Interactions at High Energies,} Fermilab, August 23-29, 1979, ed. by 
T. B. W. Kirk and H. D. I. Abarbanel (Fermi National Accelerator Laboratory, 
Batavia, IL, 1979} 
\def \hb87{{\it Proceeding of the 1987 International Symposium on Lepton and 
Photon Interactions at High Energies,} Hamburg, 1987, ed. by W. Bartel 
and R. R\"uckl (Nucl.\ Phys.\ B, Proc.\ Suppl., vol.\ 3) (North-Holland, 
Amsterdam, 1988)} 
\def \ib{{\it ibid.}~} 
\def \ibj#1#2#3{~{\bf#1} (#3) #2} 
\def \ichep72{{\it Proceedings of the XVI International Conference on High 
Energy Physics}, Chicago and Batavia, Illinois, Sept. 6 -- 13, 1972, 
edited by J. D. Jackson, A. Roberts, and R. Donaldson (Fermilab, Batavia, 
IL, 1972)} 
\def \ijmpa#1#2#3{Int.\ J.\ Mod.\ Phys.\ A {\bf#1} (#3) #2} 
\def \ite{{\it et al.}} 
\def \jhep#1#2#3{JHEP {\bf#1} (#3) #2} 
\def \jpb#1#2#3{J.\ Phys.\ B {\bf#1} (#3) #2} 
\def \jpg#1#2#3{J.\ Phys.\ G {\bf#1} (#3) #2} 
\def \mpla#1#2#3{Mod.\ Phys.\ Lett.\ A {\bf#1} (#3) #2} 
\def \nat#1#2#3{Nature {\bf#1} (#3) #2} 
\def \nc#1#2#3{Nuovo Cim.\ {\bf#1} (#3) #2} 
\def \nima#1#2#3{Nucl.\ Instr.\ Meth. A {\bf#1} (#3) #2} 
\def \npb#1#2#3{Nucl.\ Phys.\ B {\bf#1} (#3) #2} 
\def \npps#1#2#3{Nucl.\ Phys.\ Proc.\ Suppl.\ {\bf#1} (#3) #2} 
\def \npbps#1#2#3{Nucl.\ Phys.\ B Proc.\ Suppl.\ {\bf#1} (#3) #2} 
\def \PDG{Particle Data Group, D. E. Groom \ite, \epjc{15}{1}{2000}} 
\def \pisma#1#2#3#4{Pis'ma Zh.\ Eksp.\ Teor.\ Fiz.\ {\bf#1} (#3) #2 [JETP 
Lett.\ {\bf#1} (#3) #4]} 
\def \pl#1#2#3{Phys.\ Lett.\ {\bf#1} (#3) #2} 
\def \pla#1#2#3{Phys.\ Lett.\ A {\bf#1} (#3) #2} 
\def \plb#1#2#3{Phys.\ Lett.\ B {\bf#1} (#3) #2} 
\def \pr#1#2#3{Phys.\ Rev.\ {\bf#1} (#3) #2} 
\def \prc#1#2#3{Phys.\ Rev.\ C {\bf#1} (#3) #2} 
\def \prd#1#2#3{Phys.\ Rev.\ D {\bf#1} (#3) #2} 
\def \prl#1#2#3{Phys.\ Rev.\ Lett.\ {\bf#1} (#3) #2} 
\def \prp#1#2#3{Phys.\ Rep.\ {\bf#1} (#3) #2} 
\def \ptp#1#2#3{Prog.\ Theor.\ Phys.\ {\bf#1} (#3) #2} 
\def \ppn#1#2#3{Prog.\ Part.\ Nucl.\ Phys.\ {\bf#1} (#3) #2} 
\def \rmp#1#2#3{Rev.\ Mod.\ Phys.\ {\bf#1} (#3) #2} 
\def \rp#1{~~~~~\ldots\ldots{\rm rp~}{#1}~~~~~} 
\def \si90{25th International Conference on High Energy Physics, Singapore, 
Aug. 2-8, 1990} 
\def \zpc#1#2#3{Zeit.\ Phys.\ C {\bf#1} (#3) #2} 
\def \zpd#1#2#3{Zeit.\ Phys.\ D {\bf#1} (#3) #2}

\end{multicols} 
 
\end{document}